\documentclass{jfm}

\usepackage[dvipdfmx]{graphicx}
\usepackage{natbib}
\usepackage{color}
\usepackage{xcolor}
\usepackage{float}

\ifCUPmtlplainloaded \else
  \checkfont{eurm10}
  \iffontfound
    \IfFileExists{upmath.sty}
      {\typeout{^^JFound AMS Euler Roman fonts on the system,
                   using the 'upmath' package.^^J}%
       \usepackage{upmath}}
      {\typeout{^^JFound AMS Euler Roman fonts on the system, but you
                   dont seem to have the}%
       \typeout{'upmath' package installed. JFM.cls can take advantage
                 of these fonts,^^Jif you use 'upmath' package.^^J}%
      }
  \else
  \fi
\fi

\ifCUPmtlplainloaded \else
  \checkfont{msam10}
  \iffontfound
    \IfFileExists{amssymb.sty}
      {\typeout{^^JFound AMS Symbol fonts on the system, using the
                'amssymb' package.^^J}%
       \usepackage{amssymb}%

      }{}
  \fi
\fi

\ifCUPmtlplainloaded \else
  \IfFileExists{amsbsy.sty}
    {\typeout{^^JFound the 'amsbsy' package on the system, using it.^^J}%
     \usepackage{amsbsy}}
    {}
\fi

\newsavebox{\astrutbox}
\sbox{\astrutbox}{\rule[-5pt]{0pt}{20pt}}

\def\tr{\textcolor{black}}
\def\tb{\textcolor{black}}

\title[On pressure impulse of a laser-induced underwater shock wave]{On pressure impulse of a laser-induced underwater shock wave}

\author[Y. Tagawa, S. Yamamoto, K. Hayasaka, and M. Kameda]%
{Yoshiyuki Tagawa$^1$\thanks{Email address for correspondence: tagawayo@cc.tuat.ac.jp}, \ns
Shota Yamamoto$^1$,\ns
Keisuke Hayasaka$^1$ \ns
and \\Masaharu Kameda$^1$\break}

\affiliation{$^1$Department of Mechanical Systems Engineering, Tokyo University of Agriculture and Technology,
Nakacho 2-24-16 Koganei, Tokyo 184-8588, Japan\\[\affilskip]}

\pubyear{2010}
\volume{650}
\pagerange{119--126}
\date{?; revised ?; accepted ?. - To be entered by editorial office}
\begin{document}

\maketitle

\begin{abstract}
We experimentally examine a laser-induced underwater shock wave with a special attention to pressure impulse, the time integral of pressure evolution. 
Plasma formation, shock-wave expansion, and pressure in water are observed simultaneously using a combined measurement system that obtains high-resolution nanosecond-order image sequences.
These detailed measurements reveal a non-spherically-symmetric distribution of pressure peak.
In contrast, remarkably, pressure impulse is found to distribute symmetrically for a wide range of experimental parameters even when the shock waves are emitted from an elongated \tb{region}.
The structure is determined to be a collection of multiple spherical shock waves originated from point-like plasmas in \tb{the} elongated region. 
\end{abstract}



\begin{figure}
\centerline{\includegraphics[width=0.8\textwidth]{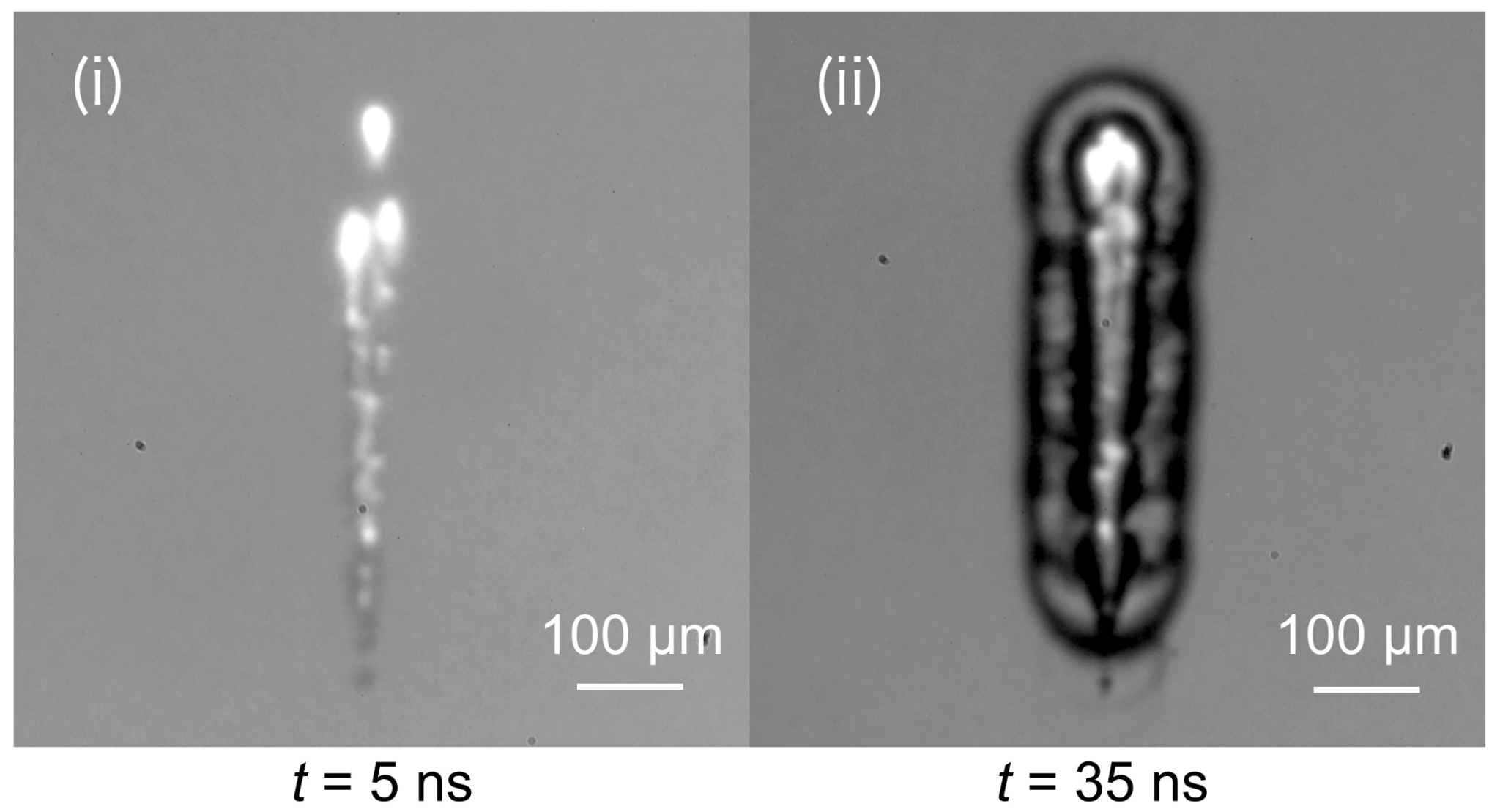}}
\caption{
{Snapshots of (i) a collection of plasmas in a conical region with 10$\times$ objective, 6.9 mJ and (ii) a collection of shock waves originated from plasmas. \tb{Time $t$ denotes the elapsed time from the start of illumination with the laser pulse.}
}}
\label{fig:Grape}       
\end{figure}

\section{Introduction} 
Underwater shock waves induced by illumination with a nanosecond laser pulse are utilized in various applications including low-invasive medical treatments \cite[]{Razvi1996, Sofer2002, Lam2002, Sankin2005, Klaseboer2007, Lee1999, Kodama2000}.
The laser-induced shock wave can trigger a sudden motion of a liquid on a free surface, which, for instance, results in generation of high-speed microjets applicable for needle-free injection devices \cite[]{Menezes2009, Thoroddsen2009, Tagawa2012, Tagawa2013, Marston2015}.

For the sudden motion of the liquid, one of the most important quantities is pressure impulse \cite[]{Batchelor1967, Cooker2006, Antkowiak2007}.
Its definition is given as:
\begin{equation}
\label{equation:impuse}
P = \int p dt,
\end{equation}
where $p$ is pressure of the liquid and $t$ is the elapsed time.
\citet{Peters2013} numerically reproduced the high-speed microjet reported by \citet{Tagawa2012} and confirmed that the pressure impulse is the key quantity for the  motion of the jet.
Thus detailed investigation for the pressure impulse of the laser-induced shock wave is of great importance.

The shock wave has been often modeled as a \textit{spherical shock}, which assumes a spherically-symmetric pressure distribution and spherical shape of the shock. 
However, some researchers have pointed out that the spherical-shock model is not applicable in certain cases \cite[]{Buzukov1969, Sankin2008, Noack1998, Vogel1996, lauterborn2013shock}.
\citet{Buzukov1969} reported that a non-spherically-symmetric bubble is observed with a series of compression waves.
\citet{Sankin2008} measured pressure peaks for a shock at various positions and determined that the peak pressure at a point in the direction perpendicular to the laser beam is more than twice as high as that in the direction of the laser.
\citet{Vogel1996, Noack1998} reported that the shape of a shock wave is not spherical due to conical plasma formation. 
The movements of the breakdown front during plasma formation in water had been intensively studied utilizing streak photographs (e.g. \cite{Docchio_1988_1, Docchio_1988_2} ). 
In our experiments we also observe a non-spherically-symmetric plasma, bubbles, and shock waves as shown in Figure~\ref{fig:Grape}. 
Figure~\ref{fig:Grape}(i) shows a collection of point-like plasmas in a conical region (like ``grapes of plasmas''). 
Figure~\ref{fig:Grape}(ii) shows another snapshot of both bubbles and shock waves. 
It confirms that \tb{for small numerical apertures ($N\!As$) of an optical system} the shock waves are not from a single conical plasma but from a collection of plasmas in a conical region. 
\tb{Note that \cite{Vogel1996_1} did show single conical plasmas.}
Despite these considerations, a common model for \textit{the pressure impulse} of laser-induced shock waves has not been developed. 

In this study, we report on experimental observations of a laser-induced shock wave with a special attention to pressure impulse.
We also propose a new model of the shock wave to rationalize the observations. 
Such an observation is, however, challenging because each phenomenon involved in generating the shock occurs within a short time; illumination with a laser pulse first triggers the emergence of plasma in water, which leads to rapid expansion of a bubble and emission of a shock wave \cite[]{Noack1999, Lauterborn2001}.
The time scale for plasma growth is in the order of nanoseconds and the shock velocity in water is approximately 1,500 m/s.
In this study plasma growth, the expansion process of the shock, and pressure in water have been \textit{simultaneously} measured using a combined measurement system, in which ultra-high-speed recording systems and pressure sensors are installed.



\section{Experimental setup and method}
\begin{figure}
\centerline{\includegraphics[width=0.8\textwidth]{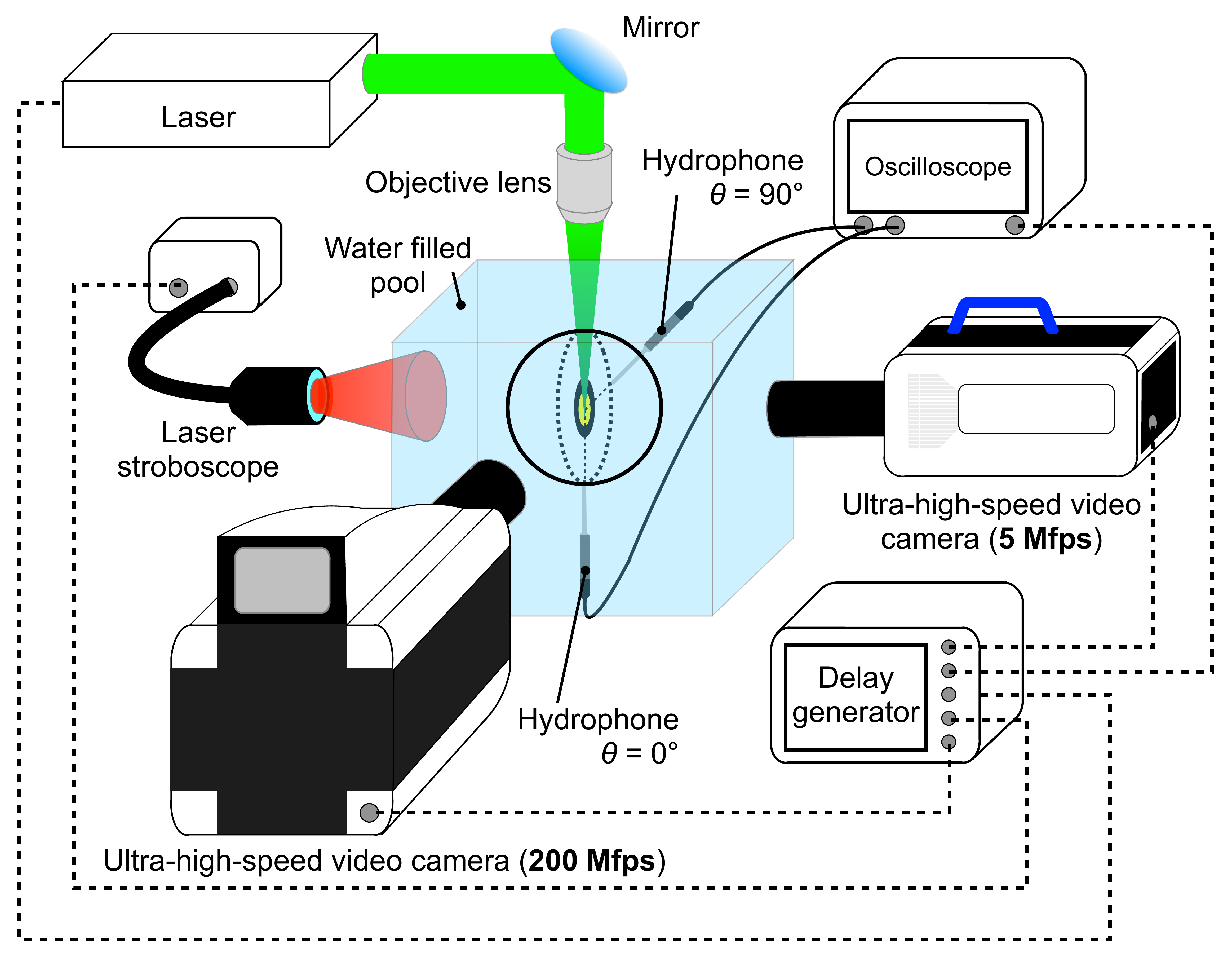}}
\caption{Measurement system consisting of two ultra-high speed cameras and two pressure sensors.
An ultra-high speed video camera records laser-induced shock waves and bubbles at up to 5$\times$10$^6$ fps with a synchronized laser stroboscope.
Plasma luminescence is captured by another ultra-high speed video camera at up to 200$\times$10$^6$ fps. 
Temporal pressure evolution is measured by two hydrophones.
One hydrophone is arranged in the direction of the laser beam ($\theta$ = $0^\circ$) at a stand-off distance of ca. 5 mm from the laser focal point, while the other hydrophone is at right angles to the hydrophone ($\theta$ = $90^\circ$) at the same stand-off distance. }
\label{fig:setup}       
\end{figure}

Figure~\ref{fig:setup} shows the combined measurement system. 
An underwater shock wave is induced by a 532 nm, 6 ns laser pulse (Nd:YAG laser Nano S PIV, Litron Lasers) focused through an objective lens to a point inside a water-filled glass container (100$\times$100$\times$450 mm).
The initial laser beam diameter is 4 mm.
Water is distilled by a water-purification system (Milli-Q Integral, Merck) at room temperature (15$\sim$20 $^\circ$C) and gas-saturated.
Its electrical conductivity is 13 M$\Omega\cdot$ cm. 
The two experimental parameters are the magnification of objective lens  (5$\times$ [$N\!A$ 0.1], 10$\times$ [$N\!A$ 0.25], 20$\times$ [$N\!A$ 0.25], MPLN series, Olympus) and the laser energy (2.6 mJ, 6.9 mJ, 12.3 mJ).
The parfocalizing distance of the objective (PFD) is 45 mm for all the microscope objectives while working distances are 20 mm, 10.6 mm, and 25 mm for 5$\times$, 10$\times$, and 20$\times$ objectives, respectively. 
Focusing angles of each microscope objective are 1 degree, 4 degrees, and 6 degrees for 5$\times$, 10$\times$, and 20$\times$ objectives\tb{, respectively}.
The \tb{diffraction-limited} focused-beam diameter $d$ is calculated by following equation (\cite{Vogel2005}),
\begin{equation}
\label{equation:spot}
d=1.22\frac{\lambda}{N\!A},
\end{equation}
where $\lambda$ (= 532 nm) is the wavelength of a laser, $N\!A$ (5$\times$ [0.10], 10$\times$ [0.25], 20$\times$ [0.25]) is the numerical aperture of microscope objective.
The calculated $d$ is 6.5 $\mu$m, 2.6 $\mu$m, and 2.6 $\mu$m with 5$\times$, 10$\times$, and 20$\times$, respectively.

The combined measurement system consists of two hydrophones (Muller Platte-Gauge, Muller) and two ultra-high-speed cameras.
One of the hydrophones is placed 5 mm away from the  focal point of the laser in the direction of the laser beam ($\theta$ = $0^\circ$ direction).
The other hydrophone is at the same distance but in the direction perpendicular to the laser beam ($\theta$ = $90^\circ$ direction).
The impulse response time (rise time of an impulsive signal) of the hydrophones (the piezoelectric PVDF type hydrophone) utilized in this study is 35-45 ns. 
The hydrophones are connected to an oscilloscope (ViewGo II DS-5554A, Iwatsu) for recording hydrophones' signals. 
The sampling frequency of the oscilloscope is 2 GHz, temporal resolution of 0.5 ns. 
The oscilloscope digitalizes pressure value every 0.03 MPa. 
One of the cameras is an ultra-high-speed camera (Imacon 200, DRS Hadland) with up to 200$\times$10$^6$ fps (5 ns time interval) and a 1,200$\times$980 pixel array to record plasma formation and shock waves in the near field. 
The other camera is another ultra-high-speed video camera (Kirana, Specialized Imaging) with up to 5$\times$10$^6$ fps and a 924$\times$768 pixel array for imaging {shadowgraph} of shock-wave propagation.
This camera is synchronized with a laser stroboscope that operates with a pulse width of 20 ns as a back illumination source  {(SI-LUX 640, Specialized Imaging)}, the repetition rate of which is also up to 5$\times$10$^6$ Hz.
A digital delay generator (Model 575, BNC) is used to trigger the laser, the hydrophones, the cameras, and the stroboscope. 
Each measurement was repeated more than three times under the same experimental conditions.



\section{Results and discussion} 

\subsection{Observations and pressure measurement}
\label{Measurement}

\begin{figure}
\centerline{\includegraphics[width=0.8\textwidth]{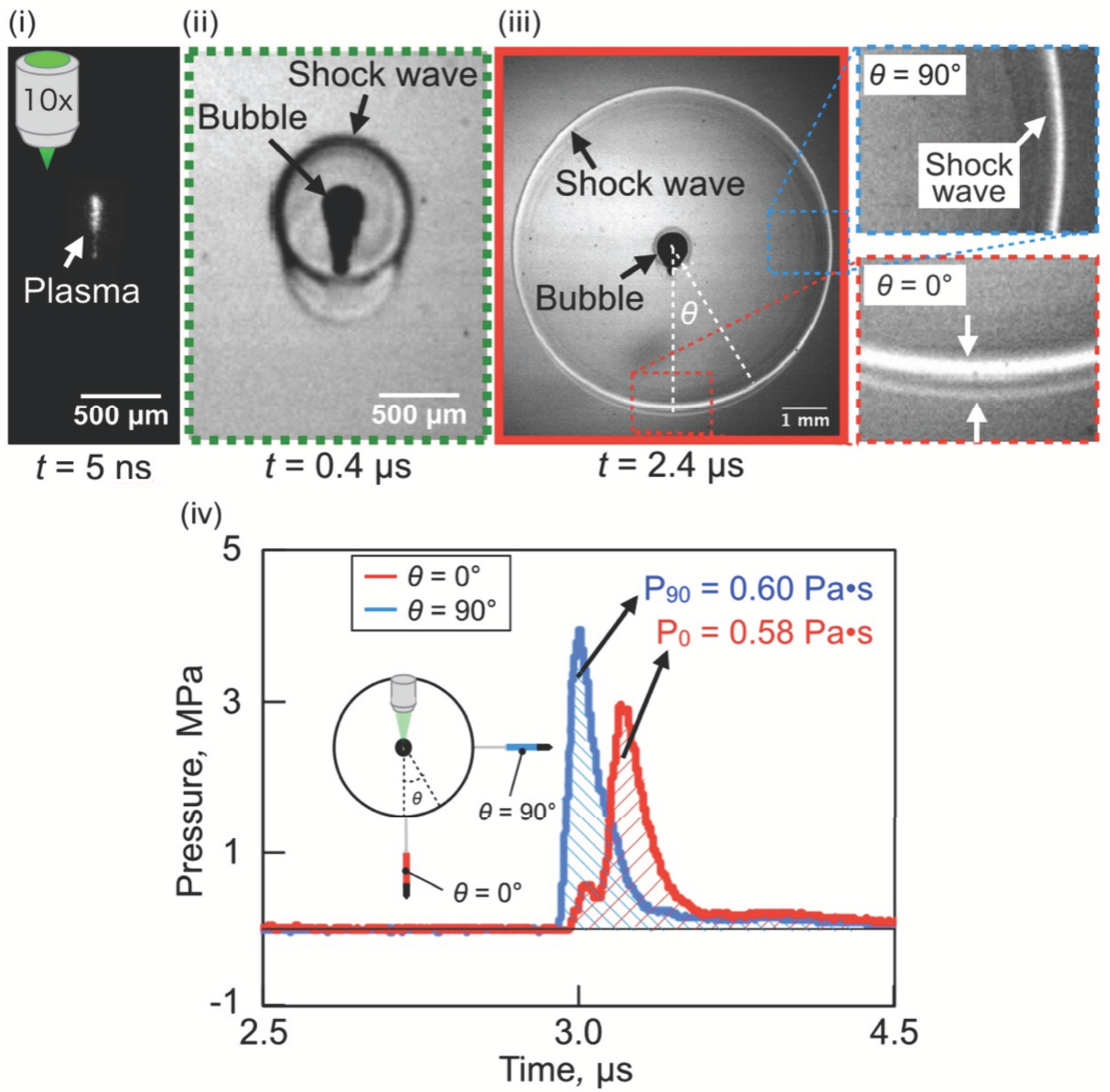}}
\caption{Measurement results for a laser-induced underwater shock wave obtained with a 10$\times$ objective lens.
(i) Plasma luminescence at $t$ = 5 ns after the laser is fired. The image is captured by an ultra-high-speed video camera at 200 Mfps. 
(ii) Shock waves and bubbles at $t$ = 0.4 $\mu$s imaged with an ultra-high speed video camera at 5 Mfps. 
(iii) Shock waves at $t$ = 2.4 $\mu$s measured with an ultra-high-speed video camera at 5 Mfps.
Enlarged images for the areas of $\theta$ = $0^\circ$ and $\theta$ = $90^\circ$ are also presented.
(iv) Shock pressure measured by the two hydrophones arranged at $\theta$ = $0^\circ$ (red line) and $\theta$ = $90^\circ$ (blue line) with respect to the laser direction.
Integrations for the pressure with respect to the elapsed time indicate pressure impulses.
}
\label{fig:results10x}       
\end{figure}


\begin{figure}
\centerline{\includegraphics[width=0.8\textwidth]{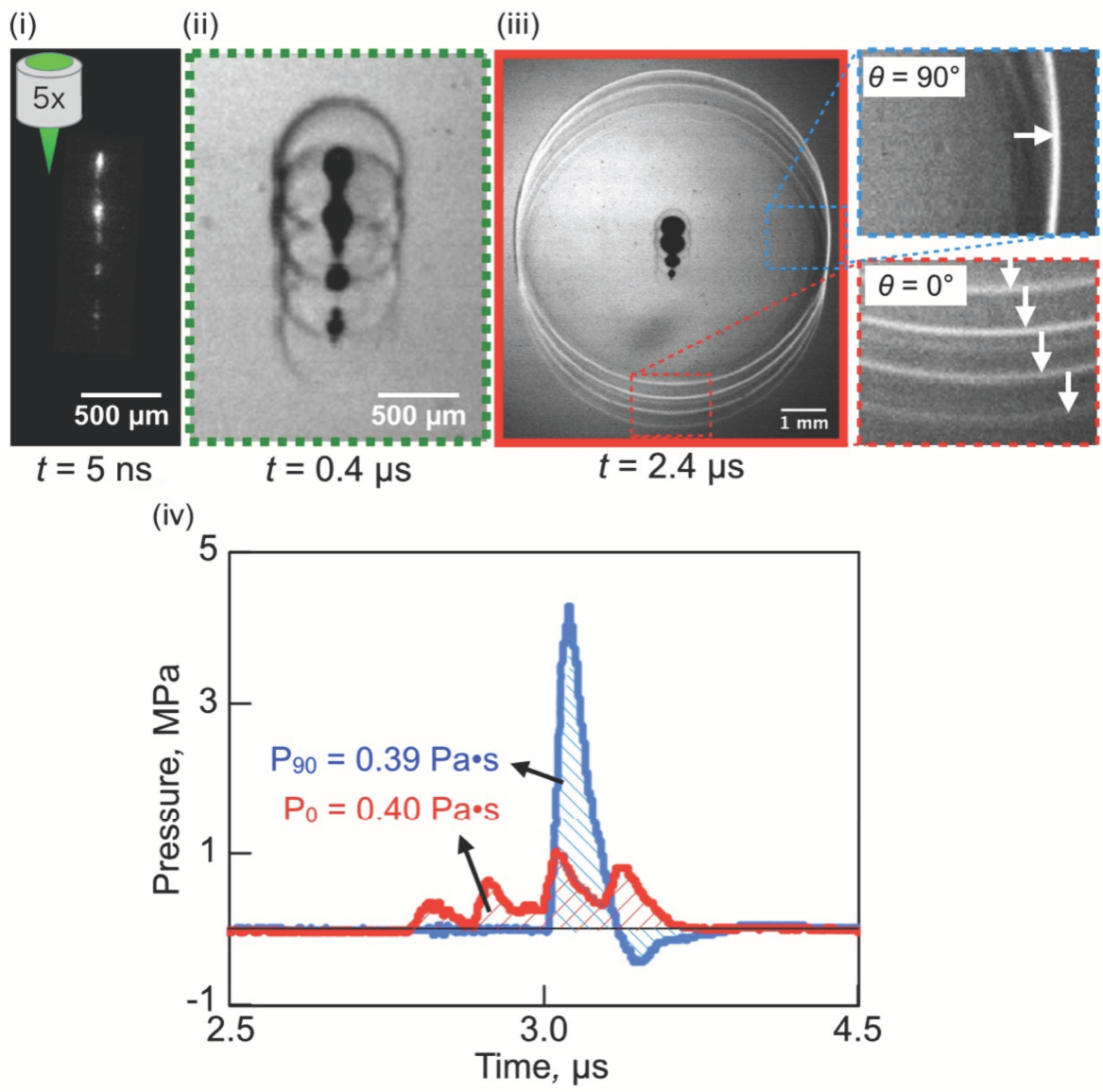}}
\caption{
{Measurement results for a laser-induced underwater shock wave obtained with a 5$\times$ objective lens.
Captions for (i)-(iv) are the same as those in Fig.~\ref{fig:results5x}
}}
\label{fig:results5x}       
\end{figure}

Figure~\ref{fig:results10x} (in which $t$ denotes the elapsed time from the start of illumination with the laser pulse) shows the measurement results obtained with the 10$\times$ objective lens.
Figure~\ref{fig:results10x}(i) shows a snapshot of the plasma luminescence in an elongated area, the major axis of which is in the direction of the laser beam.
The image sequence of the plasma confirms that all parts of the plasma emit strong lights within $\pm$5 ns.
A laser-induced bubble then emerges where the plasma was formed (Fig.~\ref{fig:results10x}(ii)) and its shape is also elongated in the direction of the laser beam.
At $t$ = 0.4 $\mu$s, {non-single} spherical shocks are observed.
In contrast, at $t$ = 2.4 $\mu$s, the shock could be regarded as a single spherical shock (see Fig.~\ref{fig:results10x}(iii)).
However, enlarged images for $\theta$ = $0^\circ$ and $\theta$ = $90^\circ$ (Fig.~\ref{fig:results10x}(iii) $\theta$ = $0^\circ$ and $\theta$ = $90^\circ$) display a clear difference. 
Two shock waves for $\theta$ = $0^\circ$ (Fig.~\ref{fig:results10x}(iii) $\theta$ = $0^\circ$), which is different from the single shock wave for $\theta$ = $90^\circ$ (Fig.~\ref{fig:results10x}(iii) $\theta$ = $90^\circ$).
Figure~\ref{fig:results10x}(iv) shows the temporal evolution of pressure measured with the two hydrophones placed at different positions.
There are two peaks for $\theta$ = $0^\circ$, while there is a single large peak for $\theta$ = $90^\circ$, which is approximately 1.3 times higher than that for $\theta$ = $0^\circ$.
Note that this dependence of the peak pressure on the angle $\theta$ is the same as that reported by \citet{Sankin2008}.
 {For the 5$\times$ objective lens, this trend is much more pronounced: four plasma groups and four shock waves separated from each other are evident (see Fig.~\ref{fig:results5x}).
 
\subsection{Pressure impulse} 

\begin{figure}
\centerline{\includegraphics[width=0.8\textwidth]{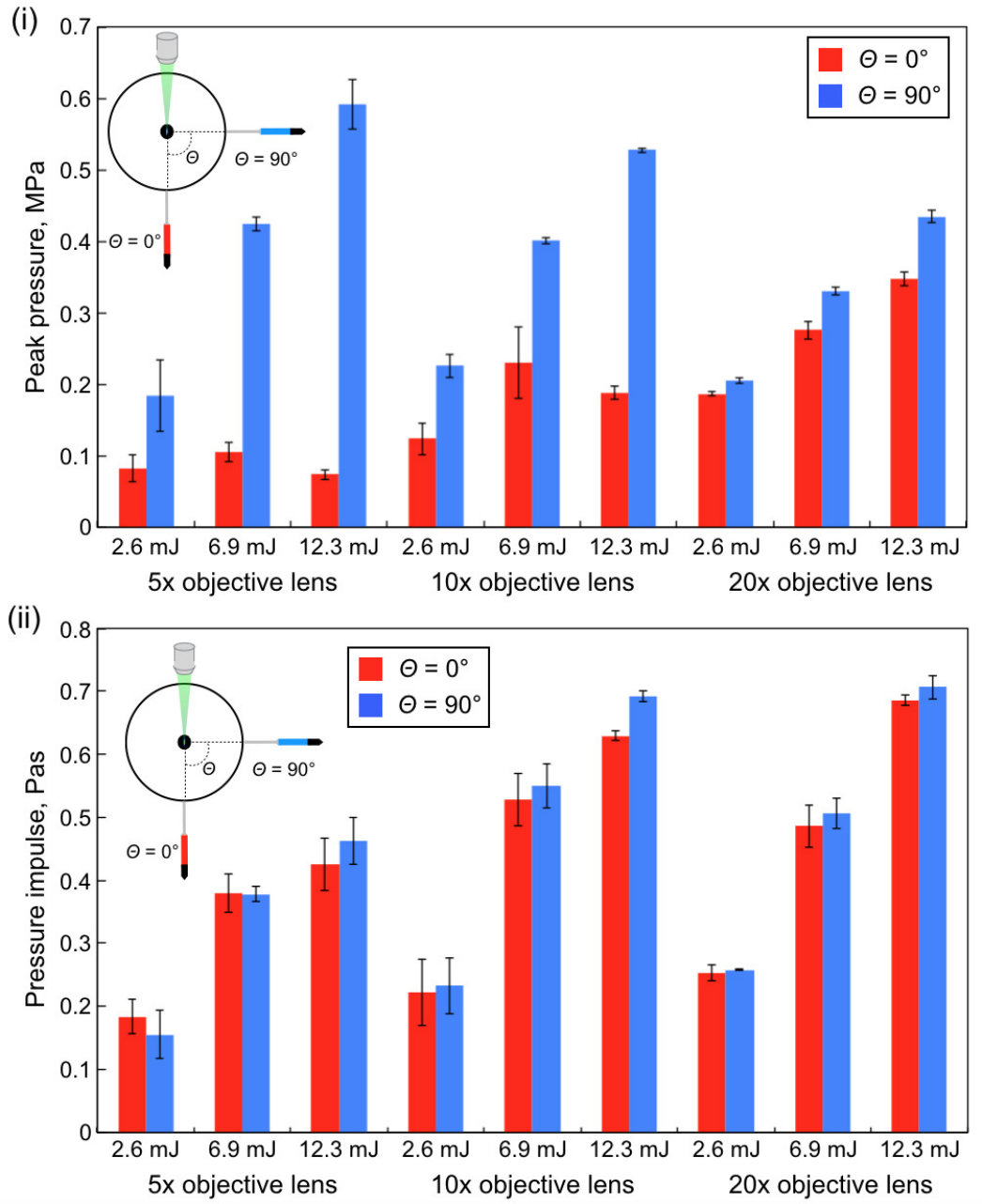}}
\caption{ {(i) \textit{Peak pressure} of a laser-induced underwater shock wave measured at $\theta$ = $0^\circ$ and $\theta$ = $90^\circ$ for all the experimental conditions. 
Each presented value is a mean for three measurements and its error bar shows the standard deviation.
(ii) \textit{Pressure impulse} of a laser-induced shock wave measured at $\theta$ = $0^\circ$ and $\theta$ = $90^\circ$ for all the experimental conditions. }
}

\label{fig:Peak_impulse}       
\end{figure}

\begin{figure}
\centerline{\includegraphics[width=0.72 \textwidth]{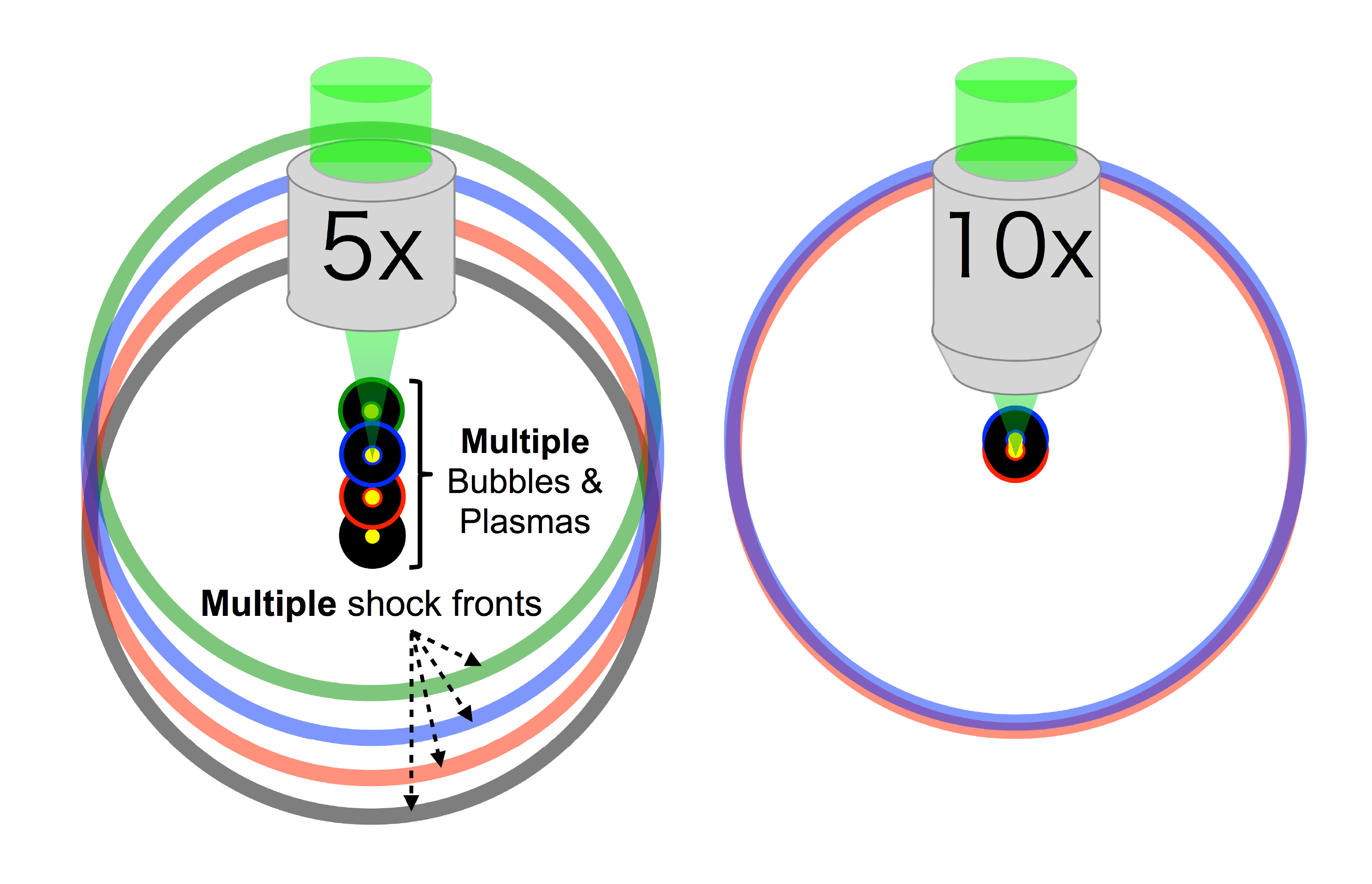}}
\caption{Schematic of a multiple structure model for a laser-induced underwater shock wave. 
Multiple plasmas emit multiple bubbles and spherical shock waves.
The shock pressure at a certain point is the sum of spherical shock pressures that reach the same point.}
\label{fig:Structure}   
\end{figure}

Here, we compute the pressure impulse for $\theta$ = $0^\circ$, $P_0$, and that for $\theta$ = $90^\circ$, $P_{90}$.
We calculate pressure impulse from $t$ = 2.5 to 4.5 $\mu$s, during which time-lags of plasma formation ($<$10 ns)  is totally covered, i.e. the time-lags do not affect the results of pressure impulse.
Integrations for the pressure with respect to the elapsed time indicate pressure impulses.
Both $P_0$ and $P_{90}$ for the shock obtained with 10$\times$ objective lens are shown in Fig.~\ref{fig:results10x}(iv) in the unit of Pa$\cdot$s.
$P_0$ is in reasonable agreement with $P_{90}$.
Furthermore, pressure impulse and  peak pressure for $\theta$ = 0$^\circ$ and 90$^\circ$ were examined for all the other experimental conditions.  
While peak pressure for $\theta$ = 0$^\circ$ and 90$^\circ$ differ significantly as shown in Fig~\ref{fig:Peak_impulse}(i), the $P_0$ was in agreement with the corresponding $P_{90}$ within the experimental uncertainty for a wide range of experimental parameters (Fig~\ref{fig:Peak_impulse}(ii))}. 
Note that the order of pressure impulse in this study is the same order of the practical use for drug delivery for cytoplasmic molecules \cite[]{Kodama2000} and generation of microjets \cite[]{Tagawa2012, Tagawa2013}.

\subsection{Structure of multiple shock waves} 
\label{Structure}

Based on the aforementioned results, we here propose a model for the structure of the laser-induced shock wave:
The shock has a multiple structure that consists of multiple spherical shock waves as depicted schematically in Fig.~\ref{fig:Structure}.
We assume that each spherical shock wave originates from the corresponding plasma formation.
In addition, since the shock wave behaves acoustically with low pressure ($\lesssim$ 100 MPa \cite[]{Vogel1996}), we could apply the \textit{superposition principle} \tb{to analyze the pressure impulse in the far-field.}
\tb{The net pressure that is produced by two or more shock waves reaching the same point is the sum of the pressure induced by the individual shock waves.}
A phenomenon that is analogous to this may be the surface wave observed after one or several stones are thrown into a quiescent pond (the so-called Huygens--Fresnel principle).
Note that the Huygens-Fresnel principle does not apply for nonlinear shock wave propagation. 

For both the 10$\times$ and 5$\times$ objective lens, this model rationalizes the observations of both pressure peaks and pressure impulse: several peaks for $\theta$ = $0^\circ$ and the single large peak for $\theta$ = $90^\circ$ while pressure impulse for $\theta$ = $0^\circ$ matches pressure impulse for $\theta$ = $90^\circ$. 
Results for a wide range of experimental parameters (Fig~\ref{fig:Peak_impulse}) may indicate the universality of the multiple structure model \tb{for optical breakdown at low or moderate $N\!A$}. 
Note that this scenario includes the well-known spherical-shock model.
It should be emphasized that, as observed with the 10$\times$ objective lens (Fig.~\ref{fig:results10x}), even if just a single plasma or a bubble is observed, the origin of the elongated plasma and bubble is expected to be multiple spots of plasma, which leads to \tb{the emergence of multiple spherical shocks}, resulting in a notable angular variation of shock pressure.
This model for the multiple shock structure could possibly rationalize the results reported in previous research.
\tb{For instance, \citet{Sankin2008} reported that a laser-induced shock wave emitted from an elongated plasma has an angular variation of pressure distribution in the far filed.}
Although the shape of the shock wave appears to be spherical, the elongated plasma may cause a multiple-structure of the shock, as observed in the present experiments (see Fig.~\ref{fig:results10x}), which would lead to a non-spherically-symmetric pressure peak of the shock.
\tb{Besides aforementioned phenomena in the far field, an anisotropy of shock wave in the near field of cylindrical plasmas had been reported (e.g. \citet[]{schoeffmann1988}), which we discuss in detail in Sec. \ref{Near field}.}

\subsection{Laser-induced plasma and bubbles} 
\label{Plasma and bubbles}
\begin{figure}
\begin{center}
\centerline{\includegraphics[width=1\textwidth]{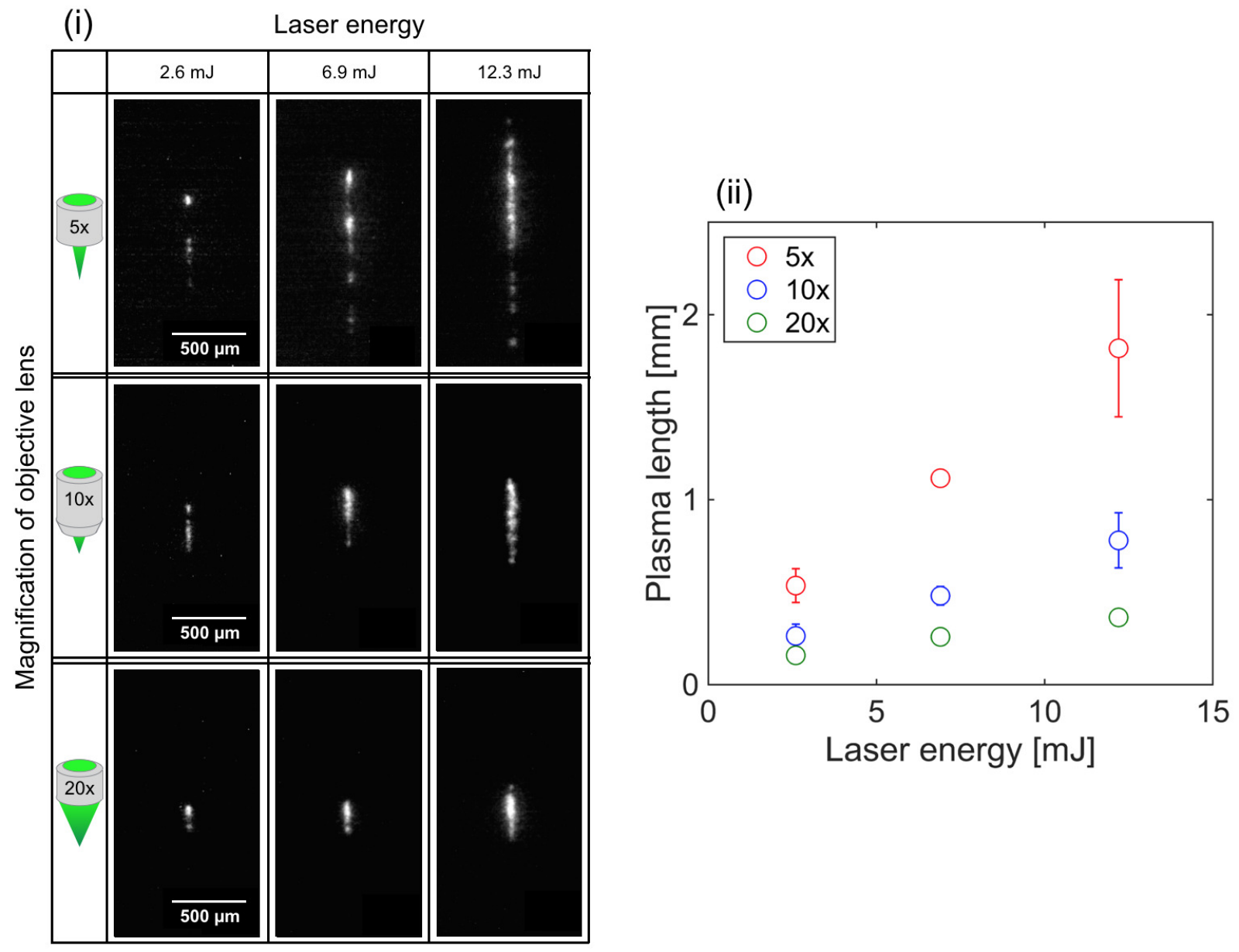}}
\end{center}
\vspace{-5mm}
\caption{(i) Plasma formation under all experimental conditions (Magnification of objective lens: 5$\times$, 10$\times$, 20$\times$. Laser energy: 2.6, 6.9, and 12.3 mJ). 
The plasma shape is the most elongated with an input energy of 12.3 mJ and the 5$\times$ objective lens, whereas it is rather spherical with an input energy of 2.6 mJ and the 20$\times$ objective lens.
(ii) Length of plasma as a function of the laser energy. The circle plot and error bar show respectively the mean and the standard deviation in 5 trials. Red, blue and green color bar represent the microscope objective of 5$\times$, 10$\times$ and 20$\times$, respectively.}
\label{fig:Plasma}       
\end{figure}

Here, attention is given to plasma formation, which is the origin of the multiple shock structure.
Figure~\ref{fig:Plasma}(i) shows the plasma luminescence under all the experimental conditions.
Figure~\ref{fig:Plasma}(ii) shows the length of the plasma region in the direction to the laser beam as a function of laser energy for 5$\times$, 10$\times$, and 20$\times$ objectives.  
The length increases with the laser energy for all objectives. 
With lower magnification, the length of the plasma region becomes longer.
The relation between an elongated plasma length (region) and focusing angle or laser pulse energy had been intensively investigated experimentally by \cite{Vogel1996_1}. 
They had found that there is a strong dependence of the elongated plasma length on the focusing angle and laser energy: 
Plasma is created in the cone angle of the laser beam proximal to the laser and becomes longer with larger laser energy. 
Our results are consistent with theirs.
Thus the multiple plasma formation is dependent on the spherical aberrations of the focusing optics, liquid impurities and the input laser energy \cite[]{Vogel1996, Vogel1996_1, Nahen1996}.

\begin{figure}
\centerline{\includegraphics[width=1\textwidth]{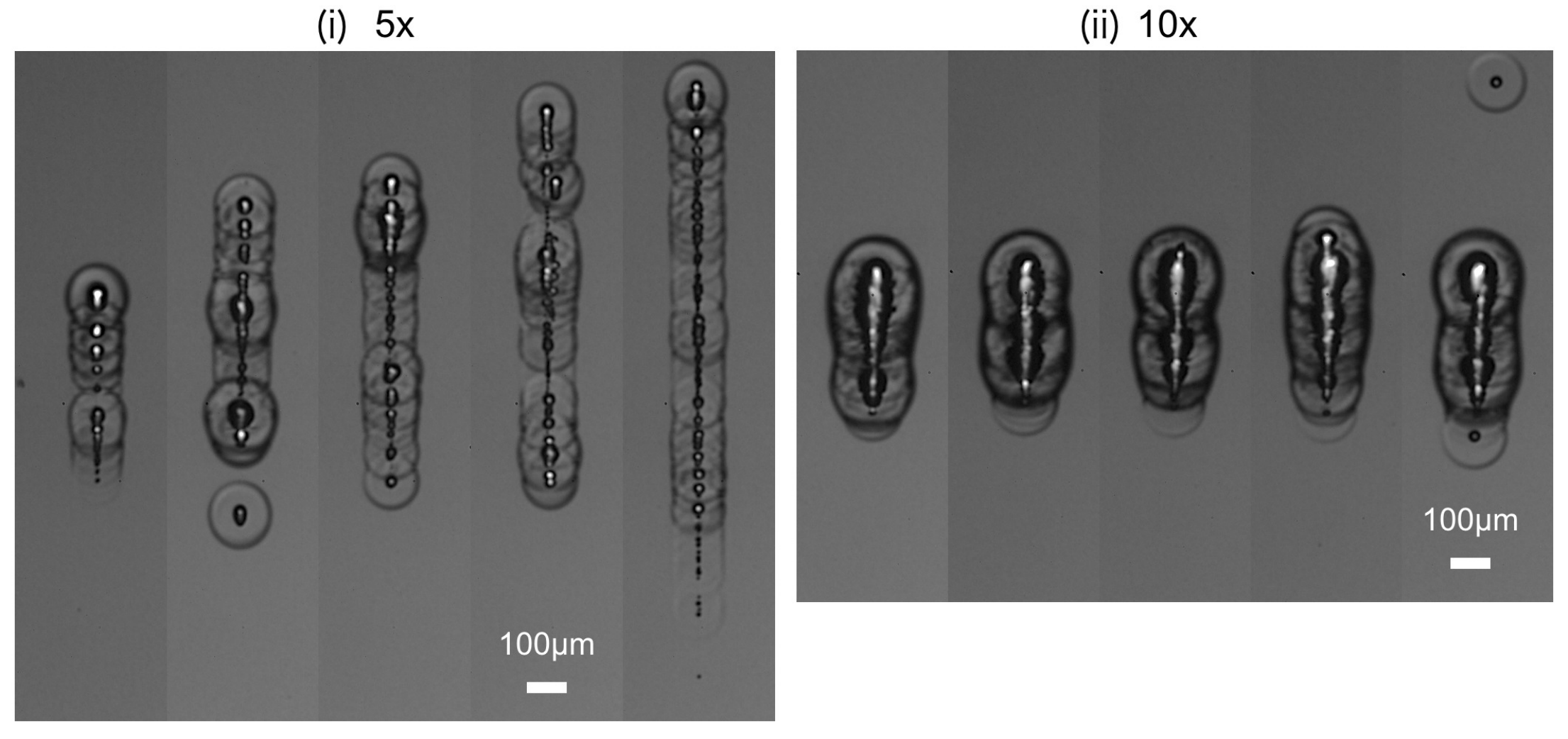}}
\caption{
{Snapshots of laser-induced bubbles and shock waves with 5$\times$, 10$\times$, 6.9 mJ, $t$ = 50 ns. These results are obtained \tb{under} the same condition.
}}
\label{fig:Longbubble}       
\end{figure}

There are two mechanisms which can lead to plasma formation: direct ionization of the medium by multiphoton absorption or avalanche ionization via inverse bremsstrahlung absorption (\cite{Vogel1996_1}).
The mechanism leading to optical breakdown (plasma formation) in this study is the avalanche ionization by heating of \tb{impurities since the position of plasma formation varies within a certain region. }
\tb{Note that impurities may provide centers for both linear and nonlinear absorption.}
\tb{If the multiphoton process in nonlinear absorption has a lower order than in pure water, impurities can trigger breakdown.}
Figure~\ref{fig:Longbubble} shows snapshots of bubbles (and shock waves) \tb{under} the same condition (the same magnification, laser energy, etc.). 
The number/position of plasma vary at every laser shot. 
If the plasma formation mechanism were purely direct ionization of the medium by multiphoton absorption, identical plasma shape should be observed for every laser shot in the same focusing optics with the same laser energy.

Note that the multiple plasma generation will not be avoided by the ``perfect'' focusing without any spherical aberrations since the plasma occurs in the region where the local energy exceeds the breakdown thresholds and thus the plasmas \tb{do} not always occur at the perfectly focused point \tb{\cite[]{vogel1999influence, Vogel2005}}. 
Therefore a focusing angle is a crucial parameter since it strongly affects local energy density.
In our experiment, there exists a strong relation between the length of the plasma region and focusing angle. 

In the case of linear sound propagation, for an extremely elongated cylindrical (or conical) source, anisotropic emission is expected with more total energy and total impulse in the directions perpendicular to the cylinder than along its axis.
However both bubbles and shock waves in the near field (e.g. Fig.~\ref{fig:Grape}) show that the source is not a single plasma but a collection of point-like plasmas in a conical region. 
Therefore, even in the case of linear sound propagation, it is not quite obvious to assume anisotropy of total energy and impulse.

\begin{figure}
\centerline{\includegraphics[width=0.5\textwidth]{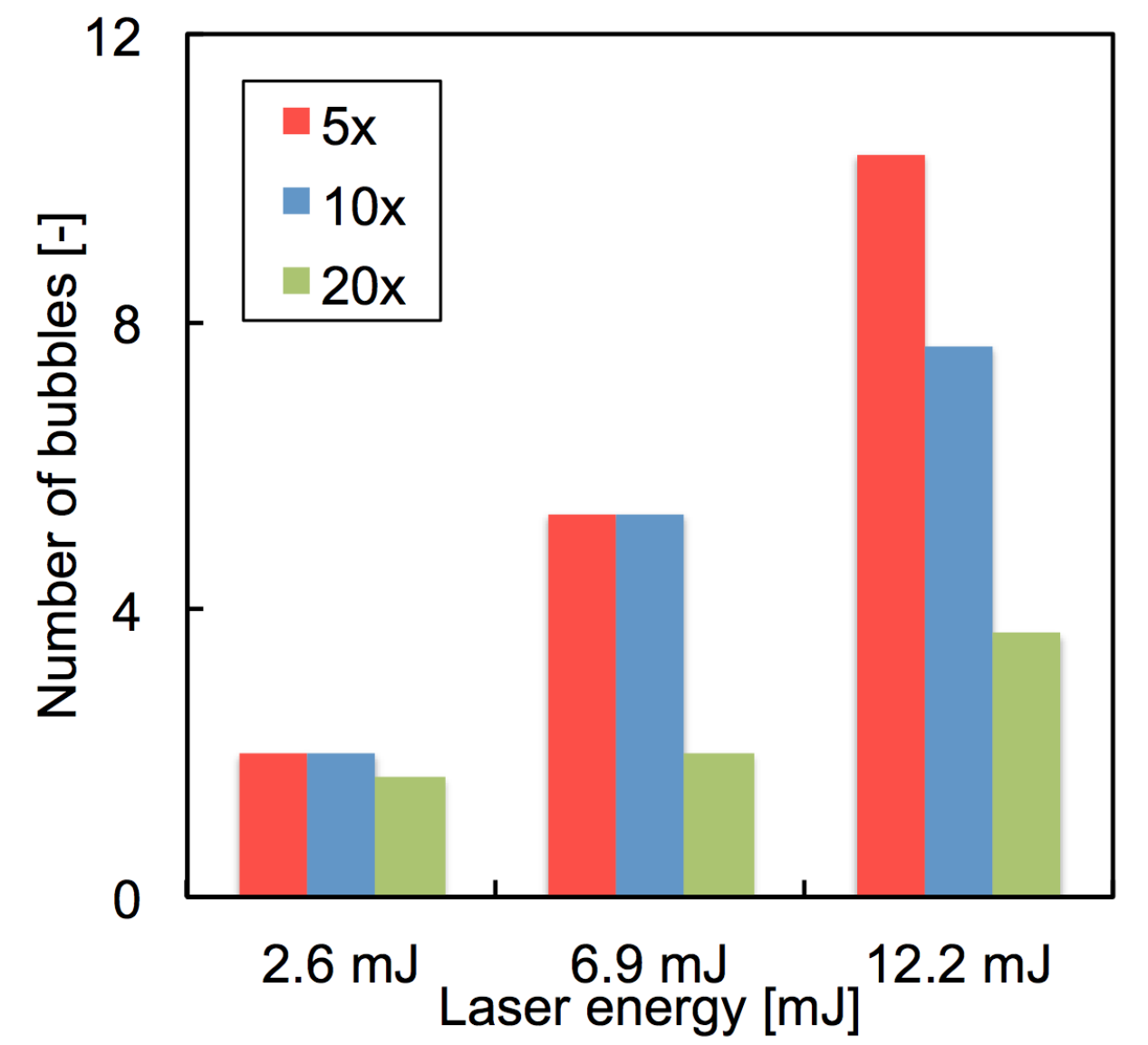}}
\caption{
{Number of bubbles of as a function of the laser energy. Each color bar \tb{represents} a result of one pulsed laser. Red, blue and green color bar represent the microscope objective of 5$\times$, 10$\times$ and 20$\times$, respectively.
}}
\label{fig:Bubbles}       
\end{figure}

\begin{figure}
\centerline{\includegraphics[width=1\textwidth]{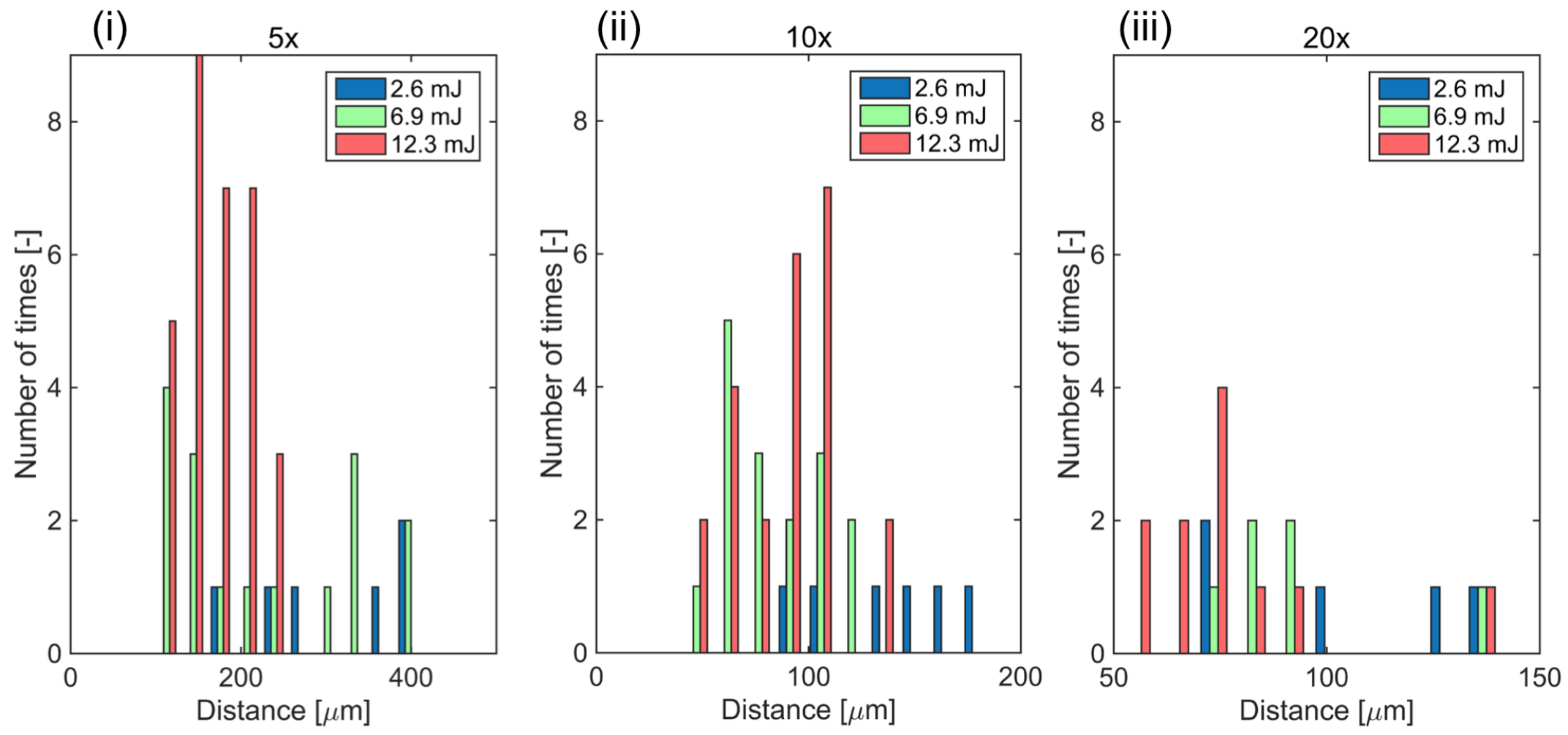}}
\caption{
{The histogram of the \tb{distances between bubble centers. 
Vertical axis is number of times that a certain distance emerges for all experiments.}
(a), (b) and (c) show the histogram of the microscope objective of 5$\times$, 10$\times$ and 20$\times$, respectively. 
The color of bar \tb{represents} the energy of a pulsed laser. The blue, green and red bar mean 2.6, 6.9 and 12.3 mJ.
}}
\label{fig:Wavelength}       
\end{figure}

The number of bubbles depends on the laser energy and the objective magnification as shown in Fig.~\ref{fig:Bubbles}. 
The number of bubbles increases with increasing laser energy and lowering the objective magnification. 
This trend is consistent with the length of plasma region as shown in Fig.~\ref{fig:Plasma}. 
\tb{We measure distances between bubble centers and count number of times that a certain distance emerges for all experimental conditions.}
Figure~\ref{fig:Wavelength}(a), (b) and (c) represent the histogram of the \tb{distances} of each bubble centers for the microscope objective of 5$\times$, 10$\times$ and 20$\times$, respectively. 
The \tb{number of times} is the sum of three trials. 
The average \tb{distance} for each condition depend on the focusing angle (microscope objective) and the laser energy.
Bubble centers are not always equidistant.

\subsection{Shock wave in the near field} 
\label{Near field}
In this section, we estimate the pressure decay in the near field to discuss energy dissipation. 
The estimation of energy dissipation for multiple plasmas and shock waves is, however, a big challenge since it is hard to disentangle the energy dissipation and nonlinear interaction in the near field. 

\begin{figure}
\centerline{\includegraphics[width=1\textwidth]{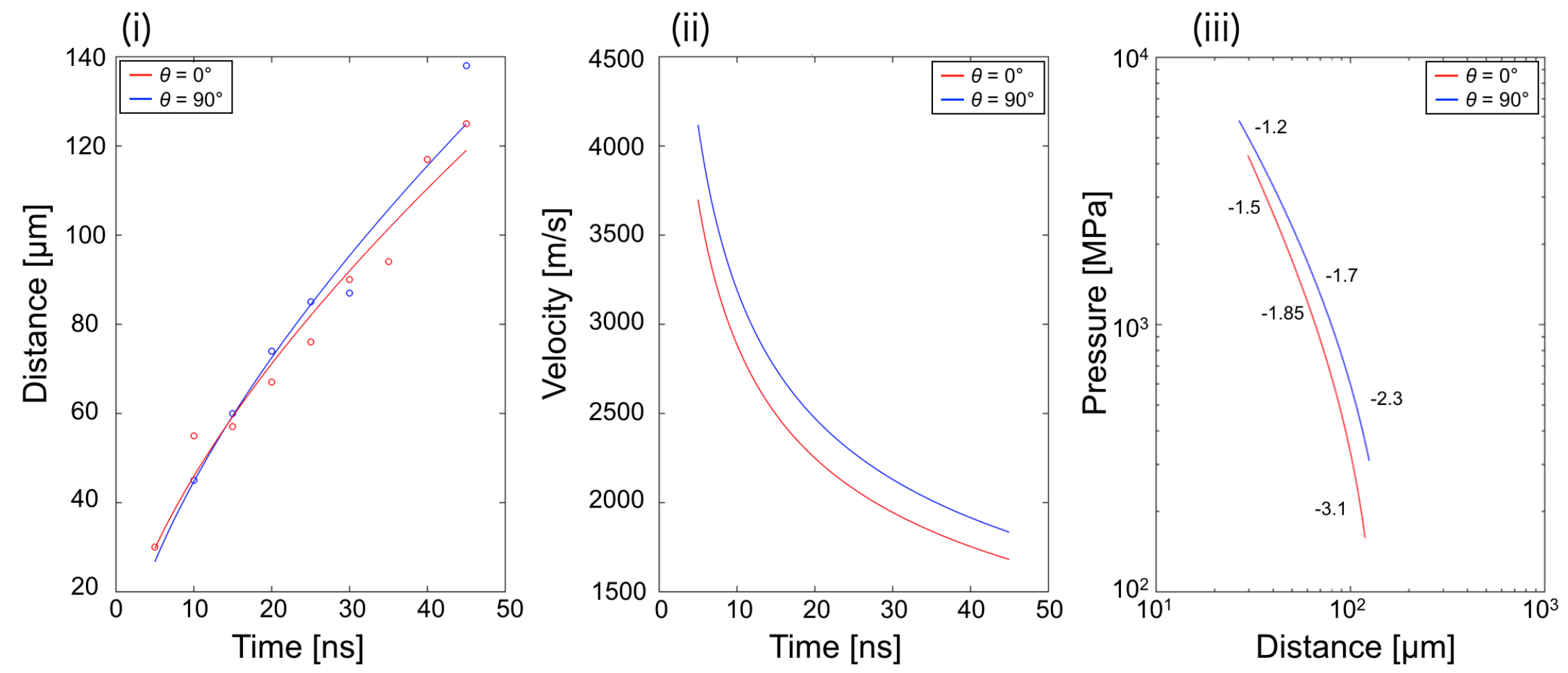}}
\caption{
{(i) The distance between plasma and a shock front vs. the elapsed time. (ii) The velocity of the shock wave vs. the elapsed time. (iii) The pressure of the shock wave vs. the distance between plasma and a shock front. The inserted numbers indicate the local slope of the corresponding curve.
}}
\label{fig:Decay}       
\end{figure}

We now use a high-speed camera for following shock wave emission from the source in the near field. 
We take snapshots of the shock wave, follow its position (Fig.~\ref{fig:Decay}(i)), derive its velocity (Fig.~\ref{fig:Decay}(ii)), and estimate the pressure (Fig.~\ref{fig:Decay}(iii)). 
To estimate pressure in a shock front $p_{s}$, we apply following equation (\cite{Vogel1996}),
\begin{equation}
\label{equation:pressure}
 p_{ s }=c_{1}\rho_{0}u_{s}({10}^{(u_{s}-c_{0}) /{c}_{2}}-1)+p_{\infty},
\end{equation}
where $\rho_{0}$ is the density of water before compression by the shock wave, $c_{0}$ is the normal sound velocity in water, $c_{1}$ = 5,190 m/s, $c_{2}$ = 25,305 m/s, and $p_{\infty}$ is the hydrostatic pressure.
The results show that, in the near field, pressure of the shock in the direction perpendicular to the laser beam is higher than that in the direction of the laser beam as shown in Fig.~\ref{fig:Decay}(iii). 
\tb{Note that \citet{Vogel1996, Noack1998} found similar pressure values close to the plasma rim as presented in the present study.}
The energy dissipation at the fronts of high-pressure shock waves is thus expected to be faster than at the low-pressure fronts \tr{since \citet{vogel1999energy} showed that the dissipation rate of acoustic energy is proportional to the pressure jump at the shock front.} 
However the pressure decay related to energy dissipation is slightly slower at the fronts of high-pressure shock wave than that at the low-pressure fronts. 
\tb{Similar results had been reported by \citet[]{schoeffmann1988, Vogel1996} that the pressure decay is not faster at the fronts of high-pressure shock wave than that at the low-pressure fronts in a certain case.}
\tb{\citet{Vogel1996}} attributed it to the formation process of the shock wave: the pressure maximum is located behind the leading edge of the pressure transient.
Our interpretation for this is that, in the near field along the direction perpendicular to the laser beam, a shock wave from one of the plasmas is followed by the other shock waves originated from the other plasmas that eventually overlap and add pressure of the shock with increasing the distance. 
This might lead to the slower pressure decay compared to the case of a single shock wave. 

\tr{
It is noteworthy that high pressures in the near field combined with nonlinear propagation and a strong anisotropy translate into the linear isotropic pressure impulse in the far field.
We discuss mechanisms of this interesting finding in this paragraph.
For a cylindrical source \citet{schoeffmann1988} showed an anisotropic shock wave emission with most energy in the 90$^\circ$ direction due to the geometrical effect. 
The anisotropic pressure jump at the initial shock front suggests that the total amount of energy dissipation is higher in the 90$^{\circ}$ direction than in the 0$^\circ$ direction (cf. \citet{vogel1999energy}).
Therefore one likely explanation for the transition is that the initial anisotropy is eroded with increasing propagation distance during nonlinear sound propagation in the transition from near field to far field, and an isotropic impulse distribution in the far field can evolve.
Another possible explanation is that nonlinear propagation in the near field has limited effects on the isotropic impulse distribution in the far field under the present experimental conditions.  
In order to address aforementioned discussions, further experimental evidence of ultra-high speed recordings would be needed.
}

\section{Conclusion and Outlook}
In order to investigate a laser-induced shock wave with focusing on pressure impulse, we constructed a measurement system consisting of a combination of ultra-high-speed cameras and pressure sensors.
Shock pressure was measured with two hydrophones arranged at $\theta$ = $0^\circ$ and $\theta$ = $90^\circ$ with respect to the laser direction, and plasma formation and shock wave expansion were simultaneously observed using two ultra-high-speed video cameras. 
The most important finding in this paper is that the distribution of pressure impulse of a shock wave is spherically symmetric \tb{(isotropic)} for a wide range of experimental parameters even when the distribution of peak pressure is non-spherically-symmetric \tb{(anisotropic)}.
We proposed a multiple structure model for the laser-induced shock wave: The laser-induced shock wave is a collection of spherical shock waves emitted from the bright spots inside the area of plasma luminescence.
The multiple structure is dependent on the plasma shape generated by illumination with the laser pulse.

To the best of the authors' knowledge, the isotropy of the pressure impulse in the far field is reported for the first time, which is of great importance for various applications. For instance, the pressure impulse in this study is in the same order of the practical use for drug delivery for cytoplasmic molecules (\cite{Kodama2000}). Other examples are low-invasive medical treatments such as drug injection and lithotripsy, for which the input energy of the laser is of the order of 10 mJ (\cite{Tagawa2012, Tagawa2013}) or even more (\cite{Menezes2009}). The laser energy in this study is in the same range: 2.6 - 12.3 mJ.
The isotropic distribution of pressure impulse may provide high degrees of freedom for the design of needle-free injection devices using high-speed microjets.
By changing the plasma shape with control parameters (magnification of the objective lens or the input laser energy), the anisotropy of the shock pressure could be controlled, which might be applicable to a variety of advanced techniques.
\\
\\
{\bf \large Acknowledgments}\\

The authors thank Shu Takagi and Yoichiro Matsumoto for the use of the Imacon 200 ultra-high-speed camera.
\tb{The authors also thank Teiichiro Ikeda for helping us to set up the ultra-high-speed camera.
This work was supported by JSPS KAKENHI Grant Number 26709007.}

\bibliographystyle{jfm.bst}
\bibliography{Reference}

\end{document}